\def\pF{p_{\text{F}}}

\def\epsilonF{\epsilon_{\text F}}

\def\be{\begin{equation}}
\def\ee{\end{equation}}
\def\bea{\begin{eqnarray}}
\def\eea{\end{eqnarray}}
\def\bse{\begin{subequations}}
\def\ese{\end{subequations}}

\documentclass[prb,twocolumn,amsmath,amssymb,eqsecnum,preprintnumbers]{revtex4}
\usepackage{graphicx}
\usepackage{dcolumn}
\usepackage{bm}
\usepackage{bbm}
\usepackage{color}
\usepackage{soul}
\begin{document}
\title{Thermal Transport and Non-Mechanical Forces in Metals}
\author{J. Amarel$^1$, D. Belitz$^{1,2}$, and T.R. Kirkpatrick$^3$}
\affiliation{$^{1}$ Department of Physics and Institute for Fundamental Science,
                    University of Oregon, Eugene, OR 97403, USA\\
                   $^{2}$ Materials Science Institute, University of Oregon, Eugene,
                    OR 97403, USA\\
                  $^{3}$ Institute for Physical Science and Technology,
                    University of Maryland, College Park,
                    MD 20742, USA
            }
\date{\today}

\begin{abstract}
We discuss contributions to the thermopower in an electron fluid. A simple argument based on
Newton's second law with the pressure gradient as the force suggests that the thermopower
is given by a thermodynamic derivative, viz., the entropy per particle, rather than being an
independent transport coefficient. The resolution is the existence of an entropic force that
results from a coupling between the mass current and the heat current in the fluid. We also
discuss and clarify some aspects of a recent paper (Phys. Rev. B {\bf 102}, 214306 (2020))
that provided a method for exactly solving electronic transport equations in the low-temperature
limit.
\end{abstract}


\maketitle

\section{Introduction}
\label{sec:I}

\subsection{Transport coefficients}
\label{subsec:I.A}

Consider the number current density ${\bm j}_n$ and the heat or entropy current density ${\bm j}_s$ in a Fermi liquid
without momentum conservation.
These currents are driven by gradients of the chemical (or electrochemical) potential $\mu$ and the temperature $T$,
and the relevant transport coefficients $L_{ij}$ are defined by the linear-response relations\cite{Callen_1985, Mahan_1981}
\bse
\label{eqs:1.1}
\bea
{\bm j}_n &=& \frac{-L_{11}}{T}\,{\bm\nabla}\mu - \frac{L_{12}}{T^2}\,{\bm\nabla}T \ ,
\label{eq:1.1a}\\
{\bm j}_s &=& \frac{-L_{21}}{T}\,{\bm\nabla}\mu - \frac{L_{22}}{T^2}\,{\bm\nabla}T \ .
\label{eq:1.1b}
\eea
\ese
Of the four transport coefficients, only three are independent, since an Onsager reciprocal relation requires $L_{21} = L_{12}$.
The independent coefficients are given by Kubo formulas\cite{Kubo_1957, Mahan_1981} that describe density-current--density-current correlations (for
$L_{11}$), density-current--heat-current correlations (for $L_{12}$), and heat-current--heat-current correlations (for $L_{22}$),
respectively. In a kinetic-theory framework they take the form of different matrix elements of the inverse collision 
operator.\cite{Dorfman_vanBeijeren_Kirkpatrick_2021, Kirkpatrick_Belitz_2022}
They determine the electrical conductivity $\sigma$, the thermopower or Seebeck coefficient ${\mathfrak S}$, and the
heat conductivity $\kappa$ via
\bse
\label{eqs:1.2}
\bea
\sigma &=& e^2 L_{11}/T\ ,
\label{eq:1.2a}\\
-e{\mathfrak S} &=& L_{12}/T\,L_{11}\ ,
\label{eq:1.2b}\\
\kappa &=& L_{22}/T^2\ ,
\label{eq:1.2c}
\eea
\ese
with $e$ the electron charge.

\subsection{A simple argument for the thermopower}
\label{subsec:I.B}

An elementary argument for the thermopower can be constructed as follows. The electron mass current density is ${\bm j}_m = m{\bm j}_n$,
with $m$ the electron mass. Let $\tau$ be the momentum relaxation time, which can be due to electron-impurity scattering, or electron-phonon
scattering, or any combination of scattering processes that do not conserve the electronic momentum. Then the equation of motion for 
${\bm j}_m$ is, by Newton's second law,
\be
\partial_t\, {\bm j}_m = \frac{-1}{\tau}\,{\bm j}_m + \frac{1}{V}\,{\bm F}\ ,
\label{eq:1.3}
\ee
with $V$ the system volume and ${\bm F}$ the total force on the electron system.\cite{current_footnote}
One contribution to the force density is the negative pressure gradient
\be
{\bm F}_p/V = -{\bm\nabla} p\ ,
\label{eq:1.4}
\ee
where $p$ is the electronic contribution to the pressure.
Let us assume for now that this is the only contribution to the force, as would be the case in a purely mechanical description of the
fluid. In steady state, $\partial_t{\bm j}_m = 0$, Eq.~(\ref{eq:1.3}) then yields
\bea
{\bm j}_n &=& -\frac{\tau}{m}\,{\bm\nabla} p
\nonumber\\
               &=&  -\frac{\tau}{m}\left(\frac{\partial p}{\partial\mu}\right)_{T,V} {\bm\nabla}\mu -  \frac{\tau}{m}\left(\frac{\partial p}{\partial T}\right)_{\mu,V} {\bm\nabla}T
\nonumber\\
               &=& -\frac{n\tau}{m}\, {\bm\nabla}\mu - \frac{s\tau}{m}\,  {\bm\nabla}T \ .             
\label{eq:1.5}
\eea               
Here $n = N/V$ and $s = S/V$ are the number density and entropy density, respectively, and in going from the second line to the third one we have
used the fact that the pressure derivatives at constant volume are just derivatives of the grand canonical potential $\Omega = -pV$. We note that
the thermodynamics of a Fermi liquid with a short-ranged interaction apply since the effective force on the electrons is the unscreened external 
force.\cite{Pines_Nozieres_1989}
Comparing with Eq.~(\ref{eq:1.1a}) we find for the electrical conductivity the Drude formula
\be
\sigma = n e^2\tau/m\ ,
\label{eq:1.6}
\ee        
and for the thermopower
\be
-e{\mathfrak S} = s/n = S/N\ .
\label{eq:1.7}
\ee
That is, according to this simple argument the thermopower is given by a thermodynamic derivative, namely, the electronic entropy per particle. 
This is because $L_{11}$ and $L_{12}$ are given by thermodynamic derivatives multiplying the same
relaxation time $\tau$, and therefore their ratio is simply a thermodynamic derivative. This is at odds
with the fact $L_{11}$ and $L_{12}$ are independent transport coefficients. 

Since $L_{11}$ and $L_{12}$ both describe the response of the same current to a driving force, there is only one
relaxation time that can appear in Eq.~(\ref{eq:1.5}). The conclusion is therefore that there must be another
contribution to the force in Eq.~(\ref{eq:1.3}) that has nothing to do with the pressure gradient. In the next
section we will use kinetic theory to elucidate the nature of this non-mechanical or entropic force.

\section{Mass transport from kinetic theory}
\label{sec:II}

\subsection{Linearized kinetic eqation}
\label{subsec:II.!}

In order to derive a kinetic theory for the mass or number current density, we recall the formalism developed
in Refs.~\onlinecite{Belitz_Kirkpatrick_2022, Kirkpatrick_Belitz_2022}. Let
\be
f_{\text{eq}}({\bm p}) = \frac{1}{e^{\xi_p/T} + 1}
\label{eq:2.1}
\ee
be the equilibrium Fermi-Dirac distribution. Here $\xi_p = \epsilon_p - \mu$, with $\mu$ the chemical
potential  and $\epsilon_p$ the equilibrium single-particle energy. Here, and in what follows, `particle'
means quasiparticle in the sense of Landau Fermi-liquid (LFL) theory.\cite{Landau_Lifshitz_IX_1991,
Baym_Pethick_1991} For simplicity we consider spinless fermions, and we keep only the first two
LFL parameters, $F_0$ and $F_1$. 
Let $f({\bm p},{\bm x},t)$ be the $\mu$-space or single-particle phase space
distribution function, consider small deviations from equilibrium,
\bse
\label{eqs:2.2}
\be
f({\bm p},{\bm x},t) = f_{\text{eq}}({\bm p}) + \delta f({\bm p},{\bm x},t)\ ,
\label{eq:2.2a}
\ee
and write $\delta f$ as
\be
\delta f({\bm p},{\bm x},t) = w({\bm p})\, \phi({\bm p},{\bm x},t)\ ,
\label{eq:2.2b}
\ee
with 
\bea
w({\bm p}) &=& -\partial f_{\text{eq}}({\bm p})/\partial{\epsilon_p} = \frac{1}{T}\,f_{\text{eq}}({\bm p}) \left[1 - f_{\text{eq}}({\bm p})\right]
\nonumber\\
&=& 
\frac{1}{4T \cosh^2(\xi_p/2T)}\ .
\label{eq:2.2c}
\eea
\ese

It is useful to define a scalar product in the space of ${\bm p}$-dependent functions that employs $w$ as a weight:
\be
\langle g({\bm p})\vert h({\bm p})\rangle = \frac{1}{V}\sum_{\bm p} w({\bm p})\,g({\bm p})\,h({\bm p})\ .
\label{eq:2.3}
\ee
In terms of this scalar product we can write density fluctuations as
\be
\delta n({\bm x},t) = \langle 1\vert\phi({\bm x},t)\rangle\ .
\label{eq:2.4}
\ee
Similarly, we can write velocity fluctuations as
\bse
\label{eqs:2.5}
\be
\delta{\bm u}({\bm x},t) = \frac{1}{nm} \langle{\bm p}\vert\phi({\bm x},t)\rangle
\label{eq:2.5a}
\ee
or, equivalently, the mass current density as
\be
{\bm j}_m({\bm x},t) = \langle{\bm p}\vert\phi({\bm x},t)\rangle\ .
\label{eq:2.5b}
\ee
\ese
Finally, temperature fluctuations, fluctuations of the entropy density $s = S/V$, and pressure fluctuations can 
be written as (see Ref.~\onlinecite{Belitz_Kirkpatrick_2022} for derivations)
\bse
\label{eqs:2.6}
\bea
\delta T({\bm x},t) &=& \frac{1}{c_V} \langle a_5({\bm p})\vert\phi({\bm p},{\bm x},t)\rangle \ ,
\label{eq:2.6a}\\
\delta s({\bm x},t) &=& \langle\epsilon_p\vert\phi({\bm p},{\bm x},t)\rangle - \mu\,\delta n({\bm x},t)\ ,
\label{eq:2.6b}\\
\delta p({\bm x},t) &=& \left(\frac{\partial p}{\partial T}\right)_{N,V} \delta T({\bm x},t) + \left(\frac{\partial p}{\partial n}\right)_{T,V} \delta n({\bm x},t)\ .
\nonumber\\
\label{eq:2.6c}
\eea
\ese
Here $c_V$ is the specific heat at constant volume, and 
\bse
\label{eqs:2.7}
\be a_5({\bm p}) = \epsilon_p - \langle\epsilon_p\vert 1\rangle/\langle 1\vert 1\rangle\ .
\label{eq:2.7a}
\ee
The functions 
\bea
a_1({\bm p})&\equiv& 1\ ,
\label{eq:2.7b}\\
a_{2,3,4}({\bm p}) &=& p_x,\,p_y,\,p_z
\label{eq:2.7c}
\eea
\ese
together with $a_5({\bm p})$ constitute the
five hydrodynamic modes. They are mutually orthogonal with respect to the scalar product defined in Eq.~(\ref{eq:2.3}),
and their normalizations are given by\cite{Belitz_Kirkpatrick_2022}
\bse
\label{eqs:2.8}
\bea
\langle 1\vert 1\rangle &=& (\partial n/\partial\mu)_{T,V}(1+F_0)\ ,
\label{eq:2.8a}\\
\langle{\bm p}\vert{\bm p}\rangle &=& 3 n m^*\ ,
\label{eq:2.8b}\\
\langle a_5({\bm p})\vert a_5({\bm p})\rangle &=& c_V T\ ,
\label{eq:2.8c}
\eea
\ese
with $m^* = m(1+F_1/3)$ the quasiparticle effective mass. 

Performing a Fourier transform in space and time, with ${\bm k}$ the wave number and $\omega$ the frequency, we can write the 
linearized kinetic equation for the $\mu$-space distribution function in the form\cite{Belitz_Kirkpatrick_2022, Kirkpatrick_Belitz_2022}
\be
\left[ -i\omega - \Lambda({\bm p}) + L_{\bm k}^{(1)}({\bm p})\right] \vert\phi({\bm p},{\bm k},\omega)\rangle = 0\ .
\label{eq:2.9}
\ee
Here $\Lambda({\bm p})$ is the collision operator, and 
\bea
L_{\bm k}^{(1)}({\bm p}) = i{\bm k}\cdot{\bm v}_p &+& \frac{F_0}{\langle 1\vert 1\rangle}\,\vert i{\bm k}\cdot{\bm v}_p\rangle\langle 1\vert
\nonumber\\
&+& \frac{F_1}{\langle{\bm p}\vert{\bm p}\rangle}\,\vert(i{\bm k}\cdot{\bm v}_p){\bm p}\rangle\cdot\langle{\bm p}\vert\qquad
\label{eq:2.10}
\eea
is a kinetic operator linear in ${\bm k}$ that comprises the streaming term and the Fermi-liquid interaction. ${\bm v}_p = {\bm p}/m^*$
is the quasiparticle velocity. We are interested in a physical situation where particle number and energy are conserved,
\bse
\label{eqs:2.11}
\be
\Lambda({\bm p})\vert 1\rangle = \Lambda({\bm p})\vert a_5({\bm p})\rangle = 0\ ,
\label{eq:2.11a}
\ee
but momentum is not,
\be
\Lambda({\bm p}) \vert{\bm p}\rangle \neq 0\ .
\label{eq:2.11b}
\ee
\ese

\subsection{Kinetic equation for the mass current}
\label{subsec:II.B}

We are interested in a mass current driven by gradients of the chemical potential, or the density, and the temperature.
Accordingly, we want to derive an effective theory that explicitly keeps the five hydrodynamic modes. To this end, we
define a projection operator
\bse
\label{eqs:2.12}
\be
{\cal P} = \sum_{\alpha=1}^5 \vert a_{\alpha}\rangle \frac{1}{\langle a_{\alpha}\vert a_{\alpha}\rangle} \langle a_{\alpha}\vert
\label{eq:2.12a}
\ee
that projects on the hydrodynamic space ${\cal L}_h$ spanned by the hydrodynamic modes, and another projection operator
\be
{\cal P}_{\perp} = \mathbbm{1} - {\cal P}
\label{eq:2.12b}
\ee
\ese
that projects onto the space ${\cal L}_{\perp}$ that is orthogonal to ${\cal L}_h$. Operating from the left with $\langle{\bm p}\vert{\cal P}$
on Eq.~(\ref{eq:2.9}), and using ${\cal P} + {\cal P}_{\perp} = \mathbbm{1}$ yields
\bse
\label{eqs:2.13}
\bea
\left(-i\omega +1/\tau_0\right){\bm j}_m &=& -\langle{\bm p}\vert L_{\bm k}^{(1)}{\cal P}\vert\phi\rangle + \langle{\bm p}\vert\Lambda{\cal P}_{\perp}\vert\phi\rangle
\nonumber\\
 && - \langle{\bm p}\vert L_{\bm k}^{(1)}{\cal P}_{\perp}\vert\phi\rangle
\label{eq:2.13a}
\eea
where
\be
\frac{1}{\tau_0} = \frac{-1}{\langle{\bm p}\vert{\bm p}\rangle}\,\langle{\bm p}\vert\Lambda({\bm p})\vert{\bm p}\rangle
\label{eq:2.13b}
\ee
\ese
is a bare relaxation rate for the mass current. Operating on Eq.~(\ref{eq:2.9}) from the left with ${\cal P}_{\perp}$,
and again using ${\cal P} + {\cal P}_{\perp} = \mathbbm{1}$, allows us to express ${\cal P}_{\perp}\vert\phi\rangle$
in terms of ${\bm j}_m$ and ${\cal P}\vert\phi\rangle$:
\be
{\cal P}_{\perp} \vert\phi\rangle = G\,{\cal P}_{\perp}\Lambda\vert{\bm p}\rangle\cdot \frac{1}{\langle{\bm p}\vert{\bm p}\rangle}\,{\bm j}_m 
     - G\,{\cal P}_{\perp} L_{\bm k}^{(1)}{\cal P}\vert\phi\rangle\ .
\label{eq:2.14}
\ee
Here we have defined a propagator
\bse
\label{eqs:2.15}
\be
G({\bm p},{\bm k},\omega) = \left(-i\omega - {\cal P}_{\perp}\Lambda({\bm p}){\cal P}_{\perp} + {\cal P}_{\perp} L_{\bm k}^{(1)}({\bm p}){\cal P}_{\perp}\right)^{-1}
\label{eq:2.15a}
\ee
For later reference we expand $G$ to linear order in the wave number ${\bm k}$:
\be
G = G_0 - G_0 {\cal P}_{\perp} L_{\bm k}^{(1)} {\cal P}_{\perp} G_0 + O({\bm k}^2)\ ,
\label{eq: 2.15b}
\ee
where
\be
G_0({\bm p},\omega) = \left(-i\omega - \Lambda_{\perp}({\bm p})\right)^{-1}
\label{eq:2.15c}
\ee
\ese
and $\Lambda_{\perp} = {\cal P}_{\perp}\Lambda{\cal P}_{\perp}$.

\subsection{Analysis of the equation for the mass current}
\label{subsec:II.C}

Consider the right-hand side of Eq.~(\ref{eq:2.13a}). To zeroth order in a gradient expansion the only contribution is
from the second term with $G_0$ substituted for $G$ in Eq.~(\ref{eq:2.14}). This term is proportional to ${\bm j}_m$,
and thus a contribution to the relaxation rate. To first order in a gradient expansion, all three terms formally contribute. 
However, the part of the third one that is formally of $O({\bm k})$ is proportional to ${\bm j}_m$, and in steady state
($\omega=0$) the mass current itself is proportional to ${\bm k}$. The third therefore is effectively of $O({\bm k}^2)$. 
The equation for the mass current to linear order in the gradients thus reads
\be
-i\omega\,{\bm j}_m = \frac{-1}{\tau}\,{\bm j}_m + {\bm f}_1 + {\bm f}_2\ ,
\label{eq:2.16}
\ee
where
\be
\frac{1}{\tau} = \frac{1}{\tau_0} - \frac{1}{\langle{\bm p}\vert{\bm p}\rangle} \langle{\bm p}\vert\Lambda G_0 {\cal P}_{\perp}\Lambda\vert{\bm p}\rangle\ .
\label{eq:2.17}
\ee
Of the two force density terms, the first one is
\bea
{\bm f}_1&=& -\langle{\bm p}\vert L_{\bm k}^{(1)}{\cal P}\vert\phi\rangle
\nonumber\\
&=& -i{\bm k} \left[\frac{n}{(\partial n/\partial\mu)_{T,V}} \delta n + \left(\frac{\partial p}{\partial T}\right)_{N,V} \delta T\right]\ ,\qquad
\label{eq:2.18}
\eea
where we have used the expression (\ref{eq:2.12a}) for the
projection operator and various of the thermodynamic identities derived in Appendix A of Ref.~\onlinecite{Belitz_Kirkpatrick_2022}.
Transforming back to real space and using general thermodynamic identities as well as Eq.~(\ref{eq:2.6c}) this can be written
\be
{\bm f}_1({\bm x},t) = -{\bm\nabla}p({\bm x},t)\ .
\label{eq:2.19}
\ee
We see that ${\bm f}_1$ is the density of the mechanical or Newtonian force ${\bm F}_p$ from Eq.~(\ref{eq:1.4}).

The second force density term is
\bse
\label{eqs:2.20}
\bea
{\bm f}_2 &=& -\langle{\bm p}\vert\Lambda G_0{\cal P}_{\perp} L_{\bm k}^{(1)}{\cal P}\vert\phi\rangle
\nonumber\\
&=& -\Big\langle{\bm p}\vert\Lambda\,\Lambda_{\perp}^{-1}\vert i({\bm k}\cdot{\bm v}_p)\psi_5^{\text{L}(0)}\Big\rangle \frac{1}{T}\,\delta T\ .
\label{eq:2.20a}
\eea
Here
\bea
\psi_5^{\text{L}(0)}({\bm p}) &=& a_5({\bm p}) - \frac{T}{n} \left(\frac{\partial p}{\partial T}\right)_{N,V}
\nonumber\\
&=& \epsilon_p - (Ts/n + \mu)
\label{eq:2.20b}
\eea
\ese
is the heat mode from Eq.~(3.16) in Ref.~\onlinecite{Belitz_Kirkpatrick_2022}, and the ket vector in Eq.~(\ref{eq:2.20a}) is
the divergence of the heat current, see Appendix B in Ref.~\onlinecite{Belitz_Kirkpatrick_2022}. The inverse projected
collision operator $\Lambda_{\perp}^{-1}$ acting on the heat current is to be interpreted as follows. 
Let $\vert x\rangle = \Lambda_{\perp}^{-1}\vert{\bm v}_p\psi_5^{\text{L}(0)}\rangle$. Then $\vert x\rangle$ is the solution of the integral equation
\be
\Lambda_{\perp}\vert x\rangle = \vert{\bm v}_p\psi_5^{\text{L}(0)}\rangle ,
\label{eq:2.21}
\ee 
with the solution made unique by the requirement $\vert x\rangle \in {\cal L}_{\perp}$.

Transforming back to real space we have 
\be
{\bm f}_2({\bm x},t) = \frac{-1}{3T} \Big\langle{\bm p}\Big\vert\Lambda({\bm p})\Lambda_{\perp}^{-1}({\bm p})\Big\vert{\bm v}_p \psi_5^{\text{L}(0)}({\bm p})\Big\rangle {\bm\nabla}T({\bm x},t)\ .
\label{eq:2.22}
\ee
Note that ${\bm f}_2$ is a pure temperature gradient and involves no density gradient. 
The prefactor is a matrix element that involves a heat current and a mass current. ${\bm f}_2$ thus results from the coupling between the
number density and the heat mode and represents a non-mechanical or entropic force. It is generically nonzero, but vanishes for simple
model collision operators where $\Lambda \Lambda_{\perp}^{-1}\vert{\bm v}_p \psi_5^{\text{L}(0)}\rangle \propto \vert{\bm v}_p \psi_5^{\text{L}(0)}\rangle$,
since the mass and heat currents are mutually orthogonal.

\subsection{Contributions to the thermopower}
\label{subsec:II.D}

By comparing the coefficients in Eqs.~(\ref{eq:1.1a}) and (\ref{eq:2.16}), with ${\bm f}_1$ and ${\bm f}_2$ from Eqs.~(\ref{eq:2.19}) and (\ref{eq:2.22}),
respectively, we can determine the Onsager coefficients $L_{11}$ and $L_{12}$ that determine the electrical conductivity $\sigma$ and the thermopower
$\mathfrak{S}$ according to Eqs.~(\ref{eqs:1.2}). For the former we obtain the Drude formula (\ref{eq:1.6}), with $\tau$ given by Eq.~(\ref{eq:2.17}).
For the latter we find
\be
-e{\mathfrak S} = \frac{S}{N} + \frac{1}{3nT\tau} \langle{\bm p}\vert\Lambda\Lambda_{\perp}^{-1}\vert{\bm v}_p \psi_5^{\text{L}(0)}\rangle\ ,
\label{eq:2.23}
\ee
which is Eq.~(\ref{eq:1.7}) augmented by a contribution from the entropic force.

\section{Examples}
\label{sec:III}

An evaluation of the entropic force for a given collision operator involves solving the integral equation (\ref{eq:2.21}). 
This is equivalent to solving the Boltzmann equation with the same collision operator. Alternatively, one can employ the hydrodynamic
theories developed in Refs.~\onlinecite{Belitz_Kirkpatrick_2022, Kirkpatrick_Belitz_2022}. In this section we use two common
scattering processes to illustrate how the entropic force contributes to the thermopower. 

\subsection{Disordered Fermi liquid}
\label{subsec:III.A}

As a simple example, consider the case of a Fermi liquid in the presence of quenched disorder. In this case, particle number
and energy are conserved, but momentum is not, as we have assumed in Eqs.~(\ref{eqs:2.11}). Hydrodynamic equations
for this problem were derived in Ref.~\onlinecite{Kirkpatrick_Belitz_2022}. Within this hydrodynamic formalism, the result
for the Onsager coefficients $L_{11}$ and $L_{12}$ is (see Eqs.~(3.22, 3.23) in Ref.~\onlinecite{Kirkpatrick_Belitz_2022})
\bse
\label{eqs:3.1}
\bea
L_{11} &=& \frac{-T}{m^2}\,\langle{\hat{\bm k}}\cdot{\bm p}\vert\Lambda^{-1}({\bm p})\vert{\hat{\bm k}}\cdot{\bm p}\rangle\ ,
\label{eq:3.1a}\\
L_{12} &=&  \frac{-T}{m^2}\,\langle{\hat{\bm k}}\cdot{\bm p}\vert\Lambda^{-1}({\bm p})\vert ({\hat{\bm k}}\cdot{\bm p})(\epsilon_p - \mu)\rangle\ .
\label{eq:3.1b}
\eea
\ese
Here the collision operator describes both electron-impurity scattering and electron-electron
scattering, and its inverse exists in this context since none of the vectors in the matrix elements are conserved
quantities. The two coefficients are given by different matrix elements of the inverse collision operator, and thus in general are
independent. However, for the simplest possible model of a constant relaxation rate,
\be
\Lambda = \frac{-1}{\tau}\,\left[\mathbbm{1} - \frac{\vert 1\rangle\langle 1\vert}{\langle 1\vert 1\rangle} - \frac{\vert a_5\rangle\langle a_5\vert}{\langle a_5\vert a_5\rangle}\right]
\label{eq:3.2}
\ee
the rate drops out of the ratio $L_{12}/L_{11}$ and performing the integrals yields 
\be
-e{\mathfrak S} = \frac{\pi^2}{2}\,T/\epsilonF\ ,
\label{eq:3.3}
\ee
which is equal to the entropy per particle of a Fermi liquid in the low-temperature limit.\cite{Baym_Pethick_1991, pressure_footnote} For this simple model collision
operator we thus recover the result (\ref{eq:1.7}) of the naive argument in Sec.~\ref{subsec:I.B}. This is consistent with
the analysis in Sec.~\ref{subsec:II.C}: With Eq.~(\ref{eq:3.2}) for the collision operator, the solution of Eq.~(\ref{eq:2.21})
is $\vert x\rangle = -\tau \vert{\bm v}_p\psi_5^{\text{L}}(0)\rangle$, and hence the entropic force, Eq.~(\ref{eq:2.22}), vanishes.

This is no longer true, even within a simple relaxation-time model, if we allow for an energy dependence of the 
relaxation time. For instance, if we replace $1/\tau$ in Eq.~(\ref{eq:3.2}) by\cite{Wilson_1954}
\be
\frac{1}{\tau({\epsilon_p})} = \frac{1}{\tau}\,\sqrt{\epsilon_p/\epsilonF}
\label{eq:3.4}
\ee
and evaluate the integrals in Eqs.~(\ref{eqs:3.1}) we obtain Wilson's result\cite{Wilson_1954}
\be
-e{\mathfrak S} = \frac{\pi^2}{3}\,T/\epsilonF\ ,
\label{eq:3.5}
\ee
consistent with the fact that now the entropic force is no longer zero.

\subsection{Electron-phonon scattering}
\label{subsec:III.B}

As another example we consider the electron-phonon scattering problem with the commonly used assumption
that the phonons remain in equilibrium.\cite{Wilson_1954, Mahan_1981}  
In Ref.~\onlinecite{Amarel_Belitz_Kirkpatrick_2020}
we provided an exact solution of integral equations for transport coefficients based on the Boltzmann equation.
However, the equations solved were not quite equivalent to the Boltzmann equation since in their derivation
various factors of the electron momentum $p$ were replaced by the Fermi momentum $\pF$. As a result
of this approximation the Onsager relation $L_{21} = L_{12}$ was violated, and the results for the
transport coefficients were different from what is obtained from the Boltzmann equation proper. However,
the solutions of the integral equations as written were exact. 
Here we discuss the changes that result from not making this approximation. 

Within the formalism of Ref.~\onlinecite{Amarel_Belitz_Kirkpatrick_2020} the Onsager coefficients $L_{11}$ and
$L_{12}$ are are given by
\bse
\label{eqs:3.6}
\bea
L_{11} &=& \frac{nT}{m}\,\langle \nu^3\vert\varphi_0\rangle\ ,
\label{eq:3.6a}\\
L_{12} &=& \frac{nT}{m}\,\langle \nu^3\vert\varphi_1\rangle\ ,
\label{eq:3.6b}
\eea
\ese
where
\be
\nu(\epsilon) = \sqrt{1 + \epsilon/\epsilonF}\ ,
\label{eq:3.7}
\ee
and the functions $\varphi_0$ and $\varphi_1$ are the solutions of integrals equations
\bse
\label{eqs:3.8}
\bea
\Lambda(\epsilon)\,\varphi_0(\epsilon) &=& -\nu(\epsilon)\ ,
\label{eq:3.8a}\\
\Lambda(\epsilon)\,\varphi_1(\epsilon) &=& -\epsilon\, \nu(\epsilon)\ ,
\label{eq:3.8b}
\eea
\ese
with $\Lambda$ a collision operator. The factors $\nu(\epsilon)$ on the right-hand sides of Eqs.~(\ref{eqs:3.8})
result from a factor $\pF/p$ that appears in the angular integrations that reduce the linearized Boltzmann equation
to a one-dimensional integral equation, see Eqs.~(C2a) and (C3) in Ref.~\onlinecite{Amarel_Belitz_Kirkpatrick_2020}.\cite{notation_footnote}
The $\nu^3$ in Eqs.~(\ref{eqs:3.6}) result from one factor of $p/\pF$ in the radial $p$-integration measure,
and one factor of $p/\pF$ from each of the two current vertices. All of these factors 
were approximated by $\nu(\epsilon) \approx 1$ in Ref.~\onlinecite{Amarel_Belitz_Kirkpatrick_2020}.

The Onsager coefficients $L_{21}$ and $L_{22}$ can also be obtained from the solutions of Eqs.~(\ref{eqs:3.8}):
\bse
\label{eqs:3.9}
\bea
L_{21} &=& \frac{nT}{m}\,\langle \epsilon\,\nu^3\vert\varphi_0\rangle\ ,
\label{eq:3.9a}\\
L_{22} &=& \frac{nT^2}{m}\,\langle \epsilon\,\nu^3\vert\varphi_1\rangle\ .
\label{eq:3.9b}
\eea
\ese

The collision operator $\Lambda$ is given by
\be
\Lambda(\epsilon) = \int du \left[K(\epsilon,u) R_{\epsilon\to u} - K_0(\epsilon,u)\right]
\label{eq:3.10}
\ee
with $R_{\epsilon\to u}\,f(\epsilon) = f(u)$ replacement operator. The kernel $K$ has three contributions,
\be
K(\epsilon,u) = K_0(\epsilon,u) - K_1(\epsilon,u) - K_2(\epsilon,u)\ .
\label{eq:3.11}
\ee
$K_0$ and $K_1$ are given by Eqs.~(2.18a) - (2.18c) in Ref.~\onlinecite{Amarel_Belitz_Kirkpatrick_2020}. 
$K_2$ gets modified by a factor of $(\pF/p)^2$ in the last line of Eq.~(C3) in that reference that also had
been approximated by $1$. This leads to
\be
K_2(\epsilon,u) = \frac{1}{2}\,\left(1 - \frac{T_1^2}{2\epsilonF^2}\right)\,\left(\frac{u-\epsilon}{T_1}\right)^2 K_0(\epsilon,u)\ ,
\label{eq:3.12}
\ee
which replaces Eq.~(2.18d) in Ref.~\onlinecite{Amarel_Belitz_Kirkpatrick_2020}. Here $T_1$ is the bosonic
energy scale that appears in the electron-phonon collision integral and is on the order of the Debye
temperature.\cite{Amarel_Belitz_Kirkpatrick_2020}

The integral equations (\ref{eqs:3.8}) can be solved exactly in the low-temperature limit by the same method as
in Ref.~\onlinecite{Amarel_Belitz_Kirkpatrick_2020}. The result is
\bse
\label{eqs:3.13}
\bea
\sigma(T\to 0) &=& \frac{n e^2}{m}\,\frac{1}{120\,\zeta(5) g_0}\,\frac{1}{1 - T_1^2/4\epsilonF^2}\,\frac{T_1^4}{T^5} 
\nonumber\\
  && \hskip 70pt + O(1/T^3)\ .
\label{eq:3.13a}\\
-e\,\mathfrak{S}(T\to 0) &=& \frac{\pi^2}{3}\,\frac{T}{\epsilonF} + O(T^3)\ .
\label{eq:3.13b}\\
\kappa(T\to 0)/T &=& \frac{n}{m}\frac{1}{g_0}\left( \eta + \frac{\pi^4/9}{120 \zeta(5)}\,
   \frac{T_1^2/\epsilonF^2}{1-T_1^2/4\epsilonF^2}\right)\frac{T_1^2}{T^3} 
   \nonumber\\
   && \hskip 70pt + O(1/T)\ .
\label{eq:3.13c}
\eea
\ese
Here $g_0$ is the electron-phonon coupling constant from Eq.~(2.6) in Ref.~\onlinecite{Amarel_Belitz_Kirkpatrick_2020},
and $\eta$ is the number from Eq.~(3.39b) in that paper. These results replace Eqs.~(3.36a), (3.37), and (3.39a),
respectively, in the same reference. The Onsager relation $L_{21} = L_{12}$ is now satisfied, and
the result for the thermopower agrees with Wilson's solution of the Boltzmann equation.\cite{Wilson_1954} We emphasize
that these results are exact solutions of the Boltzmann equation in the low-temperature limit. The result for the thermopower,
Eq.~(\ref{eq:3.13b}), is consistent with the fact that the collision operator has a complicated energy dependence and hence
the entropic force, Eq.~(\ref{eq:2.22}), does not vanish.

\section{Summary and Conclusion}
\label{sec:IV}

In summary, we have identified two physically different contributions to the thermopower in a metal. One is due to
the mechanical force on the electrons, i.e., the gradient of the pressure of the Fermi liquid. The other one
is an entropic force that arises from the mass current coupling to the heat current. This is analogous at some level to the
contributions to the sound velocity in either a classical fluid\cite{Forster_1975} or a fermionic quantum fluid.\cite{Belitz_Kirkpatrick_2022}
A purely mechanical theory would conclude that the speed of sound is given by the isothermal compressibility
of the fluid; it is the coupling to the heat mode that changes this to the adiabatic compressibility. However, an
important difference is that the thermopower is a transport coefficient, whereas the speed of sound is
a thermodynamic derivative; the only question is which derivative. 

We also have clarified some aspects of Ref.~\onlinecite{Amarel_Belitz_Kirkpatrick_2020}, which 
gave a method exactly solving electronic transport problems in the low-temperature limit. Specifically,
the integral equations solved exactly in that reference were not quite equivalent to the Boltzmann
equation due to some approximations in the procedure that transforms the Boltzmann equation into
a one-dimensional integral equation. These approximations are not necessary, and eliminating them
leads to the exact solution of the Boltzmann equation proper that is given in Sec.~\ref{subsec:III.B}.


\begin{thebibliography}{15}
\bibitem[{cur()}]{current_footnote}
\bibinfo{note}{For the scattering processes we will consider explicitly this is
  the only contribution to the current. For electrons in the presence of
  quenched disorder this is obviously true. For electron-phonon scattering in
  the common approximation that treats the phonons as remaining in equilibrium
  it also is true. More generally, the phonons will contribute to the momentum
  current, and the hydrodynamic description of the coupled electron-phonon
  problem will be more complicated.}
\bibitem[{pre()}]{pressure_footnote}
\bibinfo{note}{Here we consider only the Fermi-liquid contribution to the
  entropy and neglect contributions from, e.g., impurities or phonons.}
\end{thebibliography}

%
\end{document}